\documentclass{ecai}

\usepackage{latexsym}
\usepackage{amssymb}
\usepackage{amsmath}
\usepackage{amsthm}
\usepackage{array}
\usepackage{booktabs}
\usepackage{enumitem}
\usepackage{graphicx}
\usepackage{multirow}
\usepackage{color}
\usepackage{tcolorbox}
\usepackage{makecell} 
\usepackage{tabularx} 
\usepackage{ragged2e} 
\usepackage{bm}
\usepackage{subfiles} 
\usepackage{stfloats}
\usepackage{pdfpages}
\usepackage[font=small,labelfont=bf]{caption}

\newcolumntype{L}{>{\RaggedRight\hangafter=1\hangindent=0em}X}

\newcommand{\ul}[1]{\underline{#1}}
\newcommand{\BibTeX}{B\kern-.05em{\sc i\kern-.025em b}\kern-.08em\TeX}

\let\oldfootnote\footnote
\renewcommand\footnote[1]{%
  \oldfootnote{\setstretch{0.5}#1}
}

\begin{document}

\begin{frontmatter}


\paperid{8221}

\title{DialogGraph-LLM: Graph-Informed LLMs for End-to-End Audio Dialogue Intent Recognition}

\address[a]{South China Normal University, Guangzhou, China} 
\address[b]{Xiamen Rekey Medical Technology Co., LTD, Xiamen, China}

\author[a]{\fnms{HongYu}~\snm{Liu}\footnote{Equal contributions.}} 
\author[a]{\fnms{Junxin}~\snm{Li}\footnotemark[1]} 
\author[a]{\fnms{Changxi}~\snm{Guo}}
\author[a]{\fnms{Hao}~\snm{Chen}}
\author[a]{\fnms{Yaqian}~\snm{Huang}}
\author[a]{\fnms{Yifu}~\snm{Guo}}
\author[a]{\fnms{Huan}~\snm{Yang}\footnotemark[\ast]} 
\author[a,b]{\fnms{Lihua}~\snm{Cai}
\thanks{Correspondent authors: lee.cai@m.scnu.edu.cn;~huan.yang@m.scnu.edu.cn}}
\begin{abstract}
Recognizing speaker intent in long audio dialogues among speakers has a wide range of applications, but is a non-trivial AI task due to complex inter-dependencies in speaker utterances and scarce annotated data. To address these challenges, an end-to-end framework, namely \textbf{DialogGraph-LLM}, is proposed in the current work. 
DialogGraph-LLM combines a novel Multi-Relational Dialogue Attention Network (\emph{MR-DAN}) architecture with multimodal foundation models (e.g., Qwen2.5-Omni-7B) for direct acoustic-to-intent inference. 

An adaptive semi-supervised learning strategy is designed using LLM with a confidence-aware pseudo-label generation mechanism based on dual-threshold filtering using both global and class confidences, and an entropy-based sample selection process that prioritizes high-information unlabeled instances. 
Extensive evaluations on the proprietary \emph{MarketCalls} corpus and the publicly available \emph{MIntRec} 2.0 benchmark demonstrate \textbf{DialogGraph-LLM's} superiority over strong audio and text-driven baselines. 
The framework demonstrates strong performance and efficiency in intent recognition in real world scenario audio dialogues, proving its practical value for audio-rich domains with limited supervision. 
Our code is available at \url{https://github.com/david188888/DialogGraph-LLM}
\end{abstract}

\end{frontmatter}

\section{Introduction}\label{sec:intro}

Recognizing speaker intent (e.g., assessing target customer's intent level in service/product purchase) within spoken dialogues is crucial for effective human-computer interaction, which is crucial for a wide range of AI applications in real world scenarios. However, accurately inferring intent from audio dialogue is challenging due to complex utterance inter-dependencies~\cite{huang2024audiogpt,mahmood2023llm,liu2024emotion}, which could involve speaker changes and long-range utterance dependencies.

While text-based methods exist, raw audio offers richer paralinguistic cues critical for intent understanding~\cite{cui2024recent,waheed2024speech}. 
Traditional pipelines that treat audio dialogue as sequential input often fail to capture these signals, leading to subpar recognition performance.
A second popular cascaded pipelines typically transcribe audio speech to texts using Automatic Speech Recognition (ASR) before applying text-based Natural Language Processing (NLP) techniques for intent recognition~\cite{haghani2018audio}. This method often suffers processing delays, error propagation, and loss of rich paralinguistic information from the audio signal, hindering nuanced understanding.

Recent advancements have seen the rise of large audio-language models, which integrates powerful audio encoders with Large Language Model (LLM) backbones~\cite{huang2024audiogpt,xu2025qwen2}. These models can capture both acoustic details and semantic meaning, offering improvements over traditional methods. Yet, even these advanced models often treat dialogues as simple sequences, failing to capture complex structural relationships (e.g., speaker changes, and contextual dependencies in non-adjacent utterances) inherent in audio dialogue~\cite{shen2025long, wang2024comprehensive}. Modeling this intricate structure is critical for accurate intent recognition in realistic dialogue scenarios.

Existing work \cite{joshi2022cogmen} leverages graph-based neural networks for dialogue understanding, with utterances as nodes and relationships as edges. 
However, single type of edges that models intra-modal contextual and inter-modal complementary information was employed, which is unable to accommodate more diverse relationship types among utterances~\cite{xu2019powerfulgraphneuralnetworks}. 

To overcome the aforementioned limitations, we propose a novel Multi-Relational Dialogue Attention Network (MR-DAN), which represents a dialogue as a graph~\cite{scarselli2008graph} and employs specialized edges, including temporal progression, speaker continuity, and cross-utterance semantic connections, to model diverse relationships among utterances. 
This allows the model to capture crucial discourse structure often overlooked by standard sequential approaches.
Furthermore, we employ separate attention heads dedicated to each edge type during message passing~\cite{veličković2018graphattentionnetworks} to learn distinct aggregation functions for each relationship type. 

A major practical challenge in evaluating audio dialogue intent recognition is the relative scarcity of large-scale, publicly available datasets with reliable, fine-grained intent annotations. While valuable text-based dialogue datasets exist~\cite{li2017dailydialog,budzianowski2020multiwozlargescalemultidomain}, and multimodal benchmarks incorporating audio are emerging~\cite{zhang2024mintrec20largescalebenchmarkdataset}, acquiring extensive and accurately intent labeled audios remains a costly and labor-intensive task. 
Inspired by FreeMatch and Consistent-Teacher~\cite{wang2022freematch,wang2023consistent}, we introduce an adaptive semi-supervised strategy to more effectively utilize unlabeled data in this domain. It leverages LLM's capabilities to generate pseudo-labels for unlabeled dialogues, incorporating a dynamic, class-aware filtering mechanism to ensure high-quality knowledge transfer, effectively boosting performance with minimal supervision.

In summary, our work makes the following key contributions:
\begin{itemize}
    \item \textbf{DialogGraph-LLM Framework}: We propose DialogGraph-LLM, a novel end-to-end framework that directly processes raw audio dialogues for intent recognition.
    \item \textbf{Multi-Relational Graph Neural Network}: We propose a MR-DAN, a novel attention architecture for modeling complex utterance inter-dependencies within audio dialogues, using dedicated attention head for different relationship types.
    \item \textbf{Adaptive Semi-Supervised Learning Strategy}: We propose a novel adaptive Semi-Supervised Learning (SSL) strategy tailored for audio dialogue intent recognition. This approach utilizes LLM's generative capability for pseudo-labeling and employs dynamic, Exponentially Moving Average (EMA)-updated global and per-class confidence thresholds for robust filtering, effectively addressing the common challenge of limited labeled audio data.
\end{itemize}

\begin{figure*}[htbp]
  \centering
  \includegraphics[width=0.83\linewidth]{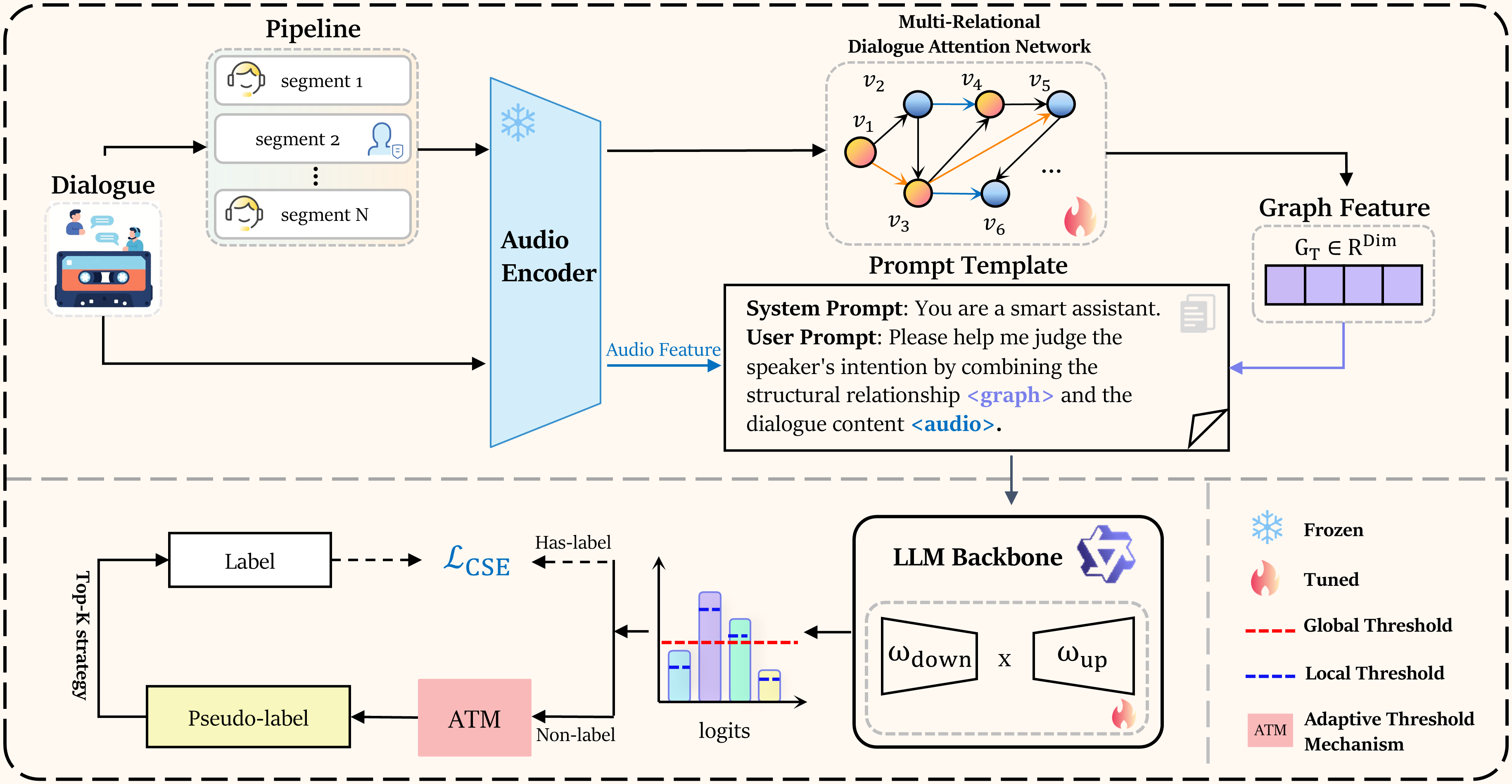}
  \caption{DialogueGraph-LLM Architecture}
  \label{fig:wide}
\end{figure*}

\section{Related Work}
\label{sec:related_work}
\subsection{Large Language Models for Audio Dialogue Understanding}

Early approaches to audio dialogue understanding often relied on cascaded pipelines, first transcribing speech via ASR~\cite{yu2016automatic} and then applying text-based models, optionally fused with basic acoustic features\cite{yoon2018multimodal, michelsanti2021overview}. These methods ignored rich paralinguistic cues that are crucial for intent recognition~\cite {waheed2024speech, zhang2023dialoguellm}.

To overcome these limitations, recent research focuses on end-to-end Audio LLMs~\cite{xu2025qwen2, rubenstein2023audiopalm, chu2023qwen,huang2024audiogpt,hurst2024gpt}. These models integrate powerful audio encoders directly with LLM backbones, enabling them to process raw audio, and jointly leverage acoustic details and linguistic contents for tasks like audio question answering and summarization~\cite{radford2023robust,baevski2020wav2vec20frameworkselfsupervised}.

Despite their capabilities in handling sequential audio, standard Audio LLMs typically lack explicit mechanisms to model complex structural relationships inherent in audio dialogues, such as speaker changes or contextual connections between non-adjacent utterances, which is crucial for nuanced intent understanding.

\subsection{Graph Neural Networks for Dialogue Modeling}   
Graph Neural Networks (GNNs) excel at modeling relational data ~\cite{scarselli2008graph, chen2019graphflow}, making them suitable for modeling dialogue structures, where utterances are treated as nodes and relationships as edges.

Previous research successfully applied GNNs to various text-based dialogue tasks, including emotion recognition~\cite{hu2021mmgcn, joshi2022cogmen} and dialogue reasoning~\cite{xing2023relational}, typically constructing graphs from textual features and metadata (e.g., speaker IDs) using architectures like GCNs~\cite{yao2019graph} or GATs~\cite{veličković2018graphattentionnetworks}.
Despite these advancements, several gaps remain, particularly in audio-based dialogue and integration with LLM. 

Firstly, recent multimodal models for dialogue understanding, including intent and emotion recognition, leveraged cross-modal interactions to enhance performance~\cite{joshi2022cogmen, li2024tracing}. However, these approaches often predominantly focus on inter-modal message passing. A critical observation, and a key inspiration for our MR-DAN, is that such a focus can lead to overlooking crucial dynamic relationships such as evolving speaker states or intricate temporal dependencies in utterances within the conversation flow. 
Secondly, while GAT introduces an attention mechanism over neighbors, previous works often use multi-head attention for a single edge type or simpler relational structures. 
This limits their ability to fully exploit specialized aggregation functions for modeling fine-grained relationships.
Lastly, effectively integrating GNN-derived structural summaries into the reasoning process of large pre-trained audio LLMs remains a unique challenge~\cite{chai2023graphllm, fan2024graph, zhang2024making}.  

\subsection{Semi-Supervised Learning for Classification Tasks}
The significant costs and efforts involved in annotating large-scale datasets, especially for complex tasks like audio intent recognition, motivate the use of SSL, which aims to leverage abundant unlabeled data alongside limited labeled data to improve model performance and generalization~\cite{reddy2018semi}. Pseudo-labeling~\cite{reddy2018semi}, where a model generates labels for unlabeled data based on high-confidence predictions, is a prominent SSL technique that is often combined with consistency regularization strategies~\cite{xie2020unsuperviseddataaugmentationconsistency,sohn2020fixmatch}.

A key limitation in many standard pseudo-labeling methods such as FixMatch~\cite{sohn2020fixmatch} is their reliance on a fixed confidence threshold, which often leads to subpar performance with class imbalance, and fails to adapt to the model's evolving confidence during training. It could also result in noisy pseudo-labels or discarding valuable samples~\cite{sohn2020fixmatch,wang2022freematch}. This highlights the need for adaptive thresholding mechanisms to improve SSL robustness and effectiveness, especially for complex models and datasets.


\begin{figure*}[htbp]
  \centering
  \includegraphics[width=0.83\textwidth]{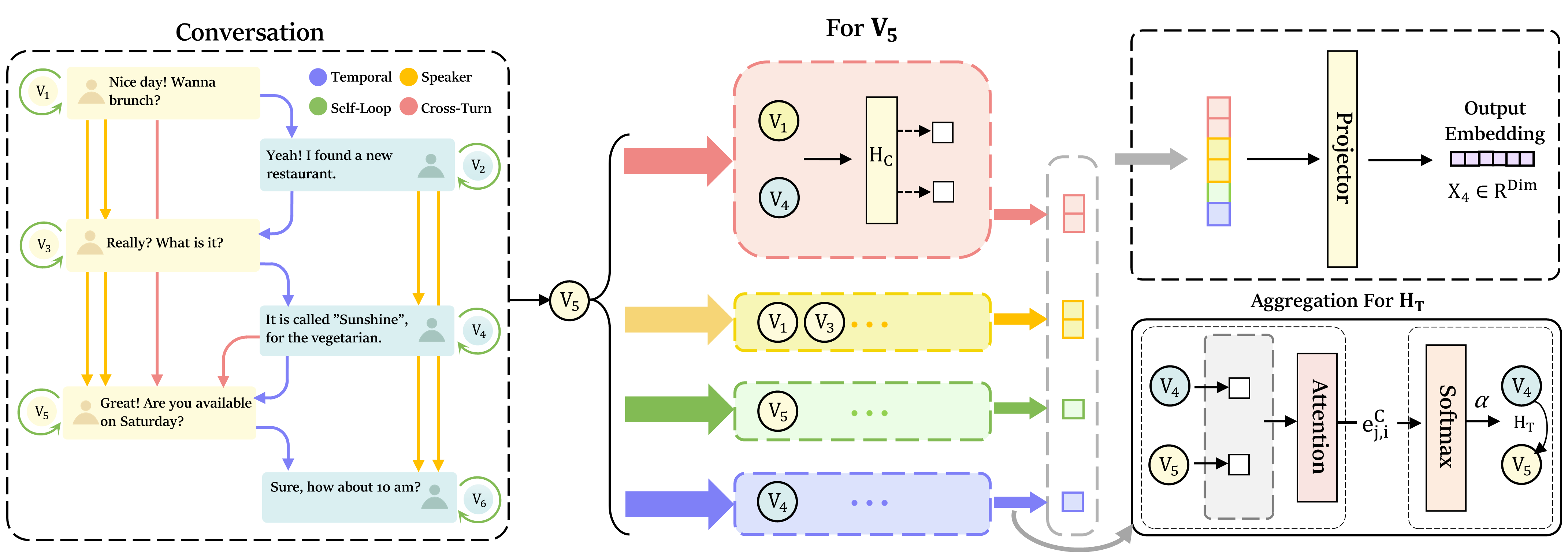}
  \caption{An example of conversation modeled by MR-DAN(left) and detailed aggregation process for $v_5$(right)}
  \label{fig:graph}
\end{figure*}

\section{Methodology}\label{sec:method}

\subsection{Problem Formulation}
\label{ssec:problem_formulation}

For an end-to-end audio dialogue intent recognition task, we denote an audio dialogue dataset as \( \{A_i, Y_i\}_{i=1}^N \), where \( A_i \) is an audio recording, and \( Y_i \) is the corresponding intent label from a predefined set.
Each audio recording \( A_i \) is first segmented into a sequence of \( l_i \) utterances, potentially associated with speaker $s_{i,j} \in S_i =\{s_{i,1}, s_{i,2}, \ldots, s_{i,l_i}\}$ for $j = 1,2,\dots,l_i$ :
\[
A_i = \{a_{i,1}, a_{i,2}, \ldots, a_{i,l_i}\}
\]
We define two primary processing functions:
\begin{enumerate}
    \item \textbf{Feature Extraction \( \Phi \):} This process maps an audio input (either the original dialogue or a segment) to a feature representation. It is applied separately to the original dialogue and each segment:
    \begin{itemize}
        \item Audio Dialogue Feature: \( G_i = \Phi(A_i) \)
        \item Audio Segment Features: \( H_i = \{h_{i,1}, h_{i,2}, \ldots, h_{i,l_i}\} \), where \( h_{i,j} = \Phi(a_{i,j}) \) for \( j = 1, \ldots, l_i \).
    \end{itemize}
    \item \textbf{Relational Structure Encoding \( \Gamma \):} This function models the structural dependencies within the dialogue based on the sequence of local utterance features \( H_i \) and speaker information \( S_i \), producing a dialogue structure representation \( R_i \):
    \[
    R_i = \Gamma(H_i, S_i)
    \]
\end{enumerate}
The goal is to develop and train a LLM model \( \mathcal{M} \) that effectively integrates both the dialogue-level audio representations \( G_i \) and the dialogue structure representation \( R_i \) to predict the overall intent \( \hat{Y}_i \):
\[
\hat{Y}_i = \mathcal{M}(G_i, R_i)
\]
The model \( \mathcal{M} \) must learn the mapping \( (H_i, R_i) \mapsto Y_i \) by leveraging the combined information derived from \( \Phi \) and \( \Gamma \).

\subsection{Audio Processing and Feature Extraction}
\label{ssec:audio_processing}
Given a raw dialogue audio recording \( A_i \), we first apply speaker diarization using techniques like binaural isolation and speaker segmentation. This process extracts a sequence of individual utterance segments \( \{a_{i,1}, a_{i,2}, \ldots, a_{i,l_i}\} \) from \( A_i \), determines their boundaries (timestamps), and assigns corresponding speaker labels, forming the speaker information set \( S_i = \{s_{i,1}, s_{i,2}, \ldots, s_{i,l_i}\} \).

For feature extraction, each original audio dialogue, along with its segmented utterance audios, is input into the pre-trained Qwen2.5-Omni audio encoder~\cite{xu2025qwen2}. The final hidden state output from the encoder is used as the primary acoustic and semantic representation for each audio and its segmented utterances, denoted as $G_i$ and $h_i$, respectively. This encoder, trained on large-scale data, provides rich representations capturing both acoustic properties (like tone, pitch) and semantic context derived from the speech content itself. The generated high-dimentional features, are suitable for downstream processing by both the MR-DAN and the LLM backbone.

\subsection{Multi-Relational Dialogue Attention Network}
\label{ssec:graph_network}
To capture the complex relational structure within a dialogue, we propose the \textbf{Multi-Relational Dialogue Attention Network}. This network explicitly models different types of relationships among utterances.

\subsubsection{Node Feature Initialization}
\label{sec:mrdan_input}
Each utterance $i$ in the dialogue is represented as a node in the graph. The initial node feature $x_i$ combines the utterance's own representation $h_i$ with information about the speaker $s_{i,j}$.
Speaker information is incorporated using learnable speaker embeddings $E_s \in \mathbb{R}^{N \times d_s}$, where $N$ is the number of unique speakers in the dialogue and $d_s$ is the speaker embedding dimension. Each speaker $s_{i,j}$ is assigned a unique vector $e_{s_j}$. These embeddings are \textbf{initialized using a zero-mean normal distribution} and learned during training. The final initial node feature $x_i$ for utterance $i$ spoken by speaker $s_i$ is formed by concatenating the utterance representation $h_i$ and the corresponding speaker embedding $e_{s_i}$, followed by a linear projection:
\begin{equation}
  x_i = W_p \cdot [h_i ; e_{s_i}]
\end{equation}
where $[;]$ denotes concatenation, and $W_p \in \mathbb{R}^{(d_h+d_s) \times d_{model}}$ is a trainable projection matrix mapping the concatenated features to the model's hidden dimension $d_{model}$.

\subsubsection{Graph Construction}
\label{sssec:graph_constrction}
A dialogue is modeled as a directed graph $G=(V,E)$, where $V$ is the set of nodes representing individual utterances, and $E$ is the set of edges representing relationships between them.

\textbf{Node Representation:} Each node $v_i \in V$ corresponds to an utterance $u_i$. Its initial feature vector $\bm{x}_i \in \mathbb{R}^d$ is obtained by projecting the concatenation of an utterance representation $\bm{h}_i$  and a learnable speaker embedding $\bm{E}_{s_i}$ associated with the speaker $s_i$ of utterance $u_i$
\begin{equation} \label{eq:node_feat}
\mathbf{x}_i = \mathbf{W}_p \cdot [\mathbf{h}_i ; \mathbf{E}_{s_i}]
\end{equation}

\textbf{Edge Definition:} We define four types of edges to capture diverse relationships among utterances:
\begin{itemize}
    \item \textbf{Temporal Edges:} Connect consecutive utterances $(v_i, v_{i+1})$ to model the sequential flow of the dialogue.
    \item \textbf{Speaker Edges:} Link an utterance $v_i$ to the $k$ most recent preceding utterances $(v_j, v_i)$ from the same speaker ($s_j = s_i, j < i$). This captures speaker-specific context and coherence. The hyperparameter $k$ controls the context window size.
    \item \textbf{Cross-Utterance Edges:} These edges capture local sequential context and semantic similarity between utterance. An edge exists from node $j$ to node $i$ if utterance $j$ immediately precedes $i$, or if $j$ occurs before $i$ and their semantic similarity $sim(x_j, x_i)$ exceeds a threshold $\theta$ . We use \textbf{cosine similarity} for $sim(x_j, x_i)$. The threshold $\theta$ is a \textbf{learnable parameter, initialized manually} at the start of training.
    \item \textbf{Self-Loops Edges:} Edges $(v_i, v_i)$ are included for each node to ensure that the node's own features are considered during feature aggregation.
\end{itemize}   
    Let $\mathcal{N}(i)$ denotes the set of all neighbors of node $v_i$. We further define $\mathcal{N}_t(i) = \{ v_j \mid (v_j, v_i) \in \mathcal{E} \text{ and edge type is } t \}$ as the set of neighbors connected to $v_i$ specifically via edges of type $t$.

\subsubsection{Multi-Relational Dialogue Attention Network Layers}
Our core contribution is the MR-DAN layer, designed to process relational information with dedicated attention heads for different edge types. 
It partitions the total $H$ attention heads into $4$ disjoint sets, $H = H_1 \cup H_2 \cup \dots \cup H_4$, where each set $H_t$ is exclusively responsible for computing attention over neighbors connected by edges of type $t$, i.e., $v_j \in \mathcal{N}_t(i)$.

\textbf{Type-Specific Attention and Aggregation:} For a given node $v_i$, each head $h \in H_t$ computes attention scores $e_{j,i}^h$ \textit{only} for neighbors $v_j \in \mathcal{N}_t(i)$. These scores are calculated based on the projected representations of the nodes:
    \begin{equation}
        e_{j,i}^h = \frac{(\bm{W}_Q^h \bm{x}_i)^T (\bm{W}_K^h \bm{x}_j)}{\sqrt{d_k}}
        \label{eq:attention_score}
    \end{equation}
where $\bm{W}_Q^h, \bm{W}_K^h, \bm{W}_V^h \in \mathbb{R}^{d_k \times d}$ are learnable projection matrices for query, key, and value transformations specific to head $h$, and $d_k = d/H$ is the dimension per head.

The scores are normalized via softmax \textit{within} the type-specific neighborhood $\mathcal{N}_t(i)$:
    \begin{equation}
        \alpha_{j,i}^h = \text{softmax}_{k \in \mathcal{N}_t(i)} (e_{k,i}^h) = \frac{\exp(e_{j,i}^h)}{\sum_{k \in \mathcal{N}_t(i)} \exp(e_{k,i}^h)}
        \label{eq:attention_norm}
    \end{equation}

Each head $h \in H_t$ then aggregates the transformed value vectors from neighbors of type $t$:
    \begin{equation}
        \bm{z}_{i,th} = \sum_{j \in \mathcal{N}_t(i)} \alpha_{j,i}^h (\bm{W}_V^h \bm{x}_j)
        \label{eq:aggregation}
    \end{equation}

\textbf{Combining Relational Information:} The outputs from all heads dedicated to the same relation type $t$ are combined (here, via concatenation) to form a type-specific aggregated representation:
    \begin{equation}
        \bm{z}_{i,t} = \underset{h \in H_t}{\text{Concat}} (\bm{z}_{i,th})
        \label{eq:combine_heads}
    \end{equation}
This yields $T$ distinct representations $\{\bm{z}_{i,1}, \bm{z}_{i,2}, \dots, \bm{z}_{i,T}\}$, each summarizing the context specific to one relation type.

\textbf{Final Node Update:} The updated node representation $\bm{x}_i^{(l+1)}$ at layer $l+1$ integrates information from all relation types by concatenating the type-specific representations, projecting them through a final matrix $\bm{W}_O$, and applying a residual connection and layer normalization (LN):
    \begin{equation}
        \bm{x}_i^{(l+1)} = \text{LN} \left( \bm{x}_i^{(l)} + \bm{W}_O [\text{Concat}(\bm{z}_{i,1}, \bm{z}_{i,2}, \dots, \bm{z}_{i,T})] \right)
        \label{eq:node_update}
    \end{equation}

\subsubsection{Graph-Level Representation Output}
\label{subsec:graph_rep}

The graph-level dialogue representation $\bm{g}$ is obtained by pooling the final node representations:
\begin{equation}
  g = \text{MEANPOOL}(\{x_i^{(L)} \mid i=1...M\})
\end{equation}
where $M$ is the number of utterances in the dialogue.

\subsection{Integrating Graph Structure and Audio Features via Prompt Engineering}

To synthesize the structural insights derived from MR-DAN with the inherent acoustic characteristics of the dialogue, we employ Qwen2.5-Omni-7B\cite{xu2025qwen2} as the core LLM backbone. This powerful multimodal model facilitates the final intent recognition by reasoning jointly over diverse input types.

A key aspect of our approach is the integration mechanism, achieved through structured prompt engineering. We design a prompt template that combines task-specific textual instructions (Prompt) with the learned representations from preceding modules. Crucially, since the graph-level dialogue representation (GraphFeature) synthesized by MR-DAN (detailed in Section~\ref{ssec:graph_network}) and the acoustic features (AudioFeature) extracted directly from the dialogue audio (as described in Section~\ref{ssec:audio_processing}) are dense numerical vectors, they cannot be directly inserted as text.

Therefore, our prompt template utilizes designated placeholders for these feature vectors. The template is structured as follows:
\[
\texttt{Prompt\_input}^k = \texttt{Template}(\texttt{Prompt}, \texttt{<graph>},\texttt{<audio>})
\]

During inference or training, the integration works as follows: The \texttt{Prompt} is converted into standard token embeddings using the LLM's text embedding layer. The \texttt{GraphFeature} ($g$) and the relevant \texttt{AudioFeature} ($h_i$) are treated as distinct multimodal inputs. These combined embeddings are then processed through the LLM's transformer layers.

This strategy enables the LLM to perform sophisticated multimodal reasoning. It leverages its pre-trained capabilities to understand the textual instructions while simultaneously considering the explicit relational structure encoded in the GraphFeature and the rich paralinguistic information contained within the AudioFeature. By processing these integrated inputs, the LLM can generate a final prediction reflecting a comprehensive understanding of the dialogue's intent, grounded in both its semantic content and its structural and acoustic properties.

\subsection{Adaptive Semi-Supervised Learning Strategy}
\label{ssec:semi_supervised}

To leverage unlabeled dialogue data ($D_U$) alongside a limited set of labeled dialogues ($D_L$), we employ an adaptive semi-supervised learning strategy based on pseudo-labeling with confidence-based filtering. The core idea involves leveraging the model's own predictions and integrating them with an adaptive threshold adjustment strategy to iteratively generate labels for unlabeled data.

\subsubsection{Leveraging LLM Predictions for Pseudo-Labeling}
\label{sssec:llm_pseudo_labeling}

The semi-supervised approach starts by direct utilization of the powerful, pre-trained \textbf{LLM backbone} within our DialogGraph-LLM model $\mathcal{M}$ for generating initial pseudo-labels. Instead of relying on traditional teacher-student models which often require separate model training or gradual knowledge transfer, we leverage the inherent strong zero-shot or few-shot prediction capabilities of LLM.

Given an unlabeled dialogue $D_u \in \mathcal{D}_U$, our LLM backbone yields a posterior probability distribution $Q_u = \mathcal{M}(D_u)$ across the $K$ potential intent classes:
\begin{equation}
Q_u = (q_{u,1}, q_{u,2}, \ldots, q_{u,K}), \quad \text{where} \sum_{c=1}^{K} q_{u,c} = 1
\end{equation}
This distribution $Q_u$ represents the model's nuanced confidence assessment for dialogue $D_u$ belonging to each intent class $c \in \{1, \dots, K\}$.

We treat these LLM-generated predictions $Q_u$ not as final pseudo-labels, but as \textbf{candidate predictions}. These candidates are rigorously filtered and refined using our proposed \textbf{Adaptive Threshold Mechanism} (detailed in Section \ref{sssec:adaptive_threshold}) to select only high-confidence and reliable pseudo-labels $\hat{y}_u$ for augmenting the training process.

\subsubsection{Adaptive Threshold Mechanism (ATM)}
\label{sssec:adaptive_threshold}
Standard pseudo-labeling often relies on a fixed, high confidence threshold, which performs poorly under class imbalance, as minority classes rarely surpass the threshold. To overcome this, we introduce a dynamic, class-aware thresholding system based on Exponential Moving Averages (EMA) of the LLM's predictions on unlabeled data.

\begin{itemize}[leftmargin=*]
    \item \textbf{Global Confidence Threshold ($\tau$):} Tracks the overall confidence of the model's predictions. It is updated via EMA based on the maximum prediction probability for each unlabeled sample in a batch $\mathcal{B}_U$:
    \begin{equation}
    \label{eq:tau_update}
    \tau_t \leftarrow \lambda \tau_{t-1} + (1 - \lambda) \mathbb{E}_{D_u \sim \mathcal{B}_U} [\max(Q_u)]
    \end{equation}
    where $\lambda$ is the EMA decay coefficient and $t$ denotes the training step. $\tau_t$ provides a baseline confidence level.

    \item \textbf{Estimated Class Distribution ($\tilde{p}$):} Estimates the model's predicted class distribution over the unlabeled data, also updated via EMA:
    \begin{equation}
    \label{eq:p_tilde_update}
    \tilde{p}_t \leftarrow \lambda \tilde{p}_{t-1} + (1 - \lambda) \mathbb{E}_{D_u \sim \mathcal{B}_U} [Q_u]
    \end{equation}
    Here, $\tilde{p}_t \in \mathbb{R}^K$ is a vector representing the smoothed average predicted probability for each class.

    \item \textbf{Class-Specific Threshold ($\tau_c$):} Dynamically computes a specific threshold for each class $c$ based on the global threshold $\tau_t$ and the estimated class distribution $\tilde{p}_t$. This allows rarer classes (with lower $\tilde{p}_{c,t}$) to have a relatively lower threshold compared to dominant classes:
    \begin{equation}
    \label{eq:tau_c_calc}
    \tau_{c,t} = \tau_t \cdot \frac{\tilde{p}_{c,t}}{\max_{k} (\tilde{p}_{k,t}) + \delta}
    \end{equation}
    where $c = 1, \ldots, K$, and $\delta$ is a small constant for numerical stability. This mechanism ensures that pseudo-labels for less frequent classes are not systematically excluded.
\end{itemize}
Both $\tau_t$ and $\tau_{c,t}$ are clamped within $[\delta, 1-\delta]$ to maintain stability.

\subsubsection{High-Confidence Pseudo-Label Selection and Balancing}
\label{sssec:filtering_balancing}

While the adaptive thresholds $\tau_{c,t}$ help accommodate class imbalance, simply accepting any prediction above these thresholds can still admit noisy pseudo-labels, particularly when the model exhibits uncertainty across multiple classes. To ensure the selection of only highly confident and unambiguous pseudo-labels, we employ a two-stage filtering process involving a \textbf{$\Delta$-Margin strategy} followed by \textbf{Class-Balanced Top-K Selection}.

\paragraph{The $\Delta$-Margin Strategy}
This initial filter assesses both absolute and relative confidence. An unlabeled dialogue $D_u$, with its predicted probability distribution $Q_u=(q_{u,1}, \dots, q_{u,K})$, is considered a candidate for pseudo-labeling with class $c^*$ if and only if the following conditions are met:
\begin{enumerate}[label=(\alph*)]
    \item \textit{Threshold Exceedance:} The predicted probability for at least one class $c$ must exceed its corresponding class-specific threshold $\tau_{c,t}$. Let $C_{valid} = \{c \mid q_{u,c} > \tau_{c,t}\}$ be the set of valid candidate classes. This set must be non-empty ($C_{valid} \neq \emptyset$).
    \item \textit{Maximum Margin Identification:} Among the valid classes, identify the class $c^*$ that exhibits the largest positive margin between its predicted probability and its threshold:
        \begin{equation}
        c^* = \arg\max_{c \in C_{valid}} (q_{u,c} - \tau_{c,t})
        \end{equation}
    \item \textit{Margin Tolerance Check:} This maximum margin must surpass a predefined tolerance boundary $\epsilon > 0$:
        \begin{equation}
        (q_{u,c^*} - \tau_{c^*,t}) > \epsilon
        \end{equation}
\end{enumerate}
This $\Delta$-Margin strategy ensures that the selected pseudo-label $c^*$ corresponds to a prediction that is not only sufficiently confident relative to its own class threshold but also significantly more confident than predictions for other competing classes relative to *their* respective thresholds, thereby filtering out ambiguous cases. Let $\mathcal{D}_{cand}$ denote the set of candidate pairs $(D_u, c^*)$ that satisfy these $\Delta$-Margin criteria within a given training period.

\paragraph{Class-Balanced Top-K Selection}
Although the $\Delta$-Margin filter enhances label quality, the resulting set $\mathcal{D}_{cand}$ might still reflect the underlying class imbalance of the dataset. To explicitly counteract this and ensure sufficient representation from minority classes in the augmented training data, we apply a class-balanced selection mechanism to $\mathcal{D}_{cand}$.
\begin{enumerate}[label=(\alph*)]
    \item \textit{Group by Pseudo-Label:} Partition the candidate set $\mathcal{D}_{cand}$ into $K$ groups based on the assigned pseudo-label class $c^*$. Let $\mathcal{D}_{cand, c} = \{(D_u, c^*) \in \mathcal{D}_{cand} \mid c^* = c\}$.
    \item \textit{Rank within Class:} Within each non-empty group $\mathcal{D}_{cand, c}$, rank the candidate dialogues $D_u$ in descending order based on their confidence score for the pseudo-labeled class, i.e., $q_{u,c}$.
    \item \textit{Select Top-K Percent:} From each ranked group $\mathcal{D}_{cand, c}$, select the top-$k_c$ candidates, where $k_c$ is typically determined as a percentage (e.g., Top $P$\%) of the group size $|\mathcal{D}_{cand, c}|$, potentially subject to a minimum count to ensure minority classes are represented if $P$\% yields too few samples. $k_c = \max(\lfloor P \cdot |\mathcal{D}_{cand, c}| \rfloor, \min\_count)$.
\end{enumerate}
The final set of selected pseudo-labeled dialogues $\mathcal{D}_{PL} = \bigcup_{c=1}^{K} \{\text{Top-}k_c \text{ samples from } \mathcal{D}_{cand, c}\}$ constitutes the high-confidence, class-balanced pseudo-labels generated in that period. These dialogues are then effectively moved from the unlabeled pool $\mathcal{D}_U$ to augment the labeled set $\mathcal{D}_L$ for subsequent training iterations. This strategy prevents dominant classes from overwhelming the pseudo-labeling process and ensures that the model continues to learn effectively from all classes, including rare ones.

\begin{table*}[t]
  \centering
   \captionsetup{font=small, labelfont=bf, width=0.8\textwidth}
  \caption{Performance comparison on MIntRec2.0~\cite{zhang2024mintrec20largescalebenchmarkdataset} Benchmark. \textbf{Left} refers to performance evaluation conducted solely on in-scope datasets, whereas \textbf{Right} measures performance on both in-scope and out-of-scope datasets within MIntRec2.0. Bold indicates best, underlined indicates second-best per task.}
  \resizebox{0.8\textwidth}{!}{
  \label{tab:performance_comparison}
  \begin{tabular}{l|cccc|cccc}
    \toprule
    \multirow{2}{*}{Methods}  & \multicolumn{4}{c|}{\textbf{In-scope Classification}} & \multicolumn{4}{c}{\textbf{In-scope + Out-of-scope Classification}} \\
    \cmidrule(lr){2-5} \cmidrule(lr){6-9} 
    & ACC (\%) $\uparrow$ & F1 (\%) $\uparrow$ & P (\%) $\uparrow$ & R (\%) $\uparrow$ & ACC $\uparrow$ & F1-IS$\uparrow$ & F1-OS$\uparrow$ & F1$\uparrow$ \\
    \midrule
    MulT~\cite{tsai2019multimodal} (ACL 2019)   & 60.66 & 54.12 & 58.02 & 53.77 & 56.00 & 46.88 & 61.66 & 47.35 \\
    MAG-BERT~\cite{rahman2020integrating} (ACL 2020) & 60.58 & 55.17 & 57.78 & 55.10 & 56.20 & 47.52 & 62.47 & 48.00 \\
    TCL-MAP~\cite{zhou2024token} (AAAI 2024) & 61.97 & \ul{56.09} & 58.14 & 53.42 & / & / & / & / \\
    A-MESS~\cite{shen2025mess}         & \ul{62.39} & 55.91 & \ul{60.10} & \ul{55.93} & \ul{56.81} & \ul{49.03} & \ul{63.42} & \ul{49.31} \\
    \midrule 
    DialogGraph-LLM (Ours)                      & \textbf{70.91} & \textbf{66.54} & \textbf{69.12} & \textbf{64.15} & \textbf{64.28} & \textbf{57.05} & \textbf{70.63} & \textbf{58.14} \\
    \bottomrule
  \end{tabular}
  }
\end{table*}

\section{Experiments}
\label{sec:experiments}

\subsection{Datasets}
\label{ssec:datasets}

\paragraph{MarketCalls} comprises 8,770 real-world telephone recordings in Mandarin, capturing interactions between sales representatives and potential customers from various companies. For the labeled data, the tags are categorized into four groups, reflecting the user's purchase intent in descending order from high to low. 
\paragraph{MIntRec2.0}~\cite{zhang2024mintrec20largescalebenchmarkdataset} is a comprehensive benchmark dataset designed for multimodal intent recognition and out-of-scope detection in multi-party conversations. It comprises annotated utterances across text, video, and audio modalities, with our evaluation focusing on samples from the text and audio modalities.

\subsection{Baselines}
\begin{itemize}[leftmargin=*]
    \item \textbf{Large Language Model}: We choose four different LLMs, including Llama3.1-8B, GLM-4-9B-0414, Gemini1.5-Pro and Qwen2.5-Omni, as our baselines for the MarketCalls dataset. 
    \item \textbf{MAG-BERT}~\cite{rahman2020integrating}: Fine-tuned pre-trained BERT for multimodal language analysis by attaching a Multimodal Adaptation Gate (MAG).
  \item \textbf{MUIT}~\cite{tsai2019multimodal}: Uses crossmodal attention Transformers to directly fuse unaligned multimodal sequences by attending to interactions across different modalities and time steps.
  \item \textbf{TCL-MAP}~\cite{zhou2024token}: A multimodal intent recognition method featuring a Modality-Aware Prompting (MAP) module to fuse nonverbal cues into prompts, and a Token-Level Contrastive Learning (TCL) framework for refining multimodal representations.
  \item \textbf{A-MESS}~\cite{shen2025mess}: Features an Anchor-based Multimodal Embedding (A-ME) module that aligns multimodal embeddings with label descriptions generated by LLM. This alignment is achieved through the integration of semantic synchronization and triplet contrastive learning.
\end{itemize}

\subsection{Experimental Settings}
\label{ssec:settings}

We fine-tune the DialogGraph-LLM framework, based on the Qwen2.5-Omni-7B backbone, using two datasets described in Section~\ref{ssec:datasets}. The AdamW optimizer~\cite{loshchilov2019decoupledweightdecayregularization} with a weight decay of 0.01 is employed, along with a cosine learning rate schedule including a linear warmup for the first 10\% of training steps.

The primary configuration utilizes Parameter-Efficient Fine-Tuning (PEFT) via Low-Rank Adaptation (LoRA)~\cite{hu2021loralowrankadaptationlarge}, with a rank of 16, alpha set to 32, and a dropout rate of 0.05. LoRA matrices are applied to the query, key, value, and output projection layers of the backbone. Learning rates are set to 2e-5 for the base model parameters and 1.6e-4 for the LoRA parameters.

MR-DAN utilizes 16-dimensional speaker embeddings with a dropout rate of 0.1. Additionally, the speaker context window size is set to 3, the initial cross-turn threshold is 0.8, and the number of attention heads is 8. Three layers of MR-DAN are used.

For Adaptive SSL, an EMA decay of 0.95 is used to update confidence thresholds and class distributions. The initial global confidence threshold is set at 0.9, with a margin tolerance of 0.06. Every 5 epochs, pseudo-labels are generated and filtered, augmenting the labeled training set with the top 10\% highest-confidence samples per class.

\begin{table}[t] 
    \centering 
    \caption{Performance Comparison of Different Models on the MarketCalls Dataset} 
    \label{tab:model_comparison_qwen_gemini_large_gap} 
    \resizebox{1\columnwidth}{!}{%
    \begin{tabular}{@{}lcccccc@{}} 
        \toprule
        \multirow{2}{*}{Model} & \textbf{Class A} & \textbf{Class B} & \textbf{Class C} & \textbf{Class D} & \multicolumn{2}{c}{\textbf{Overall}} \\
        \cmidrule(lr){2-2} \cmidrule(lr){3-3} \cmidrule(lr){4-4} \cmidrule(lr){5-5} \cmidrule(lr){6-7} 
                             & F1 (\%) & F1 (\%) & F1 (\%) & F1 (\%) & Acc (\%) & F1 (\%) \\
        \midrule

        Llama3.1-8B     & 22.70 & 56.10 & 58.50 & 19.30 & 49.85 & 49.20 \\
        GLM-4-9B-0414   & 23.60 & 58.00 & 60.30 & 20.20 & 51.75 & 51.15 \\
        \midrule 

        Gemini1.5-Pro   & 24.50 & 60.00 & 62.20 & 21.20 & 53.60 & 53.00 \\ 
        Qwen2.5-Omni    & \ul{28.50} & \ul{72.50} & \ul{74.30} & \ul{24.80} & \ul{63.58} & \ul{63.10} \\ 
        \midrule 

        \textbf{DialogGraph-LLM} & \textbf{44.53} & \textbf{83.54} & \textbf{85.21} & \textbf{41.75} & \textbf{77.31} & \textbf{76.83} \\
        \bottomrule
    \end{tabular}%
    }
    \vspace{0.5em} 
    \parbox{\linewidth}{\footnotesize \textit{Note:} Acc denotes Accuracy, F1 denotes F1-score. All values are percentages (\%). Overall metrics are calculated by weighted average.}

\end{table}

\subsection{Results and Discussion}\label{sec:results}

\paragraph{Main result on MarketCalls}
We compare the performance of our proposed model with four similarly sized LLM using our MarketCalls dataset for evaluation. The results are reported in Table~\ref{tab:performance_comparison}. It can be seen that our model surpasses all baselines in terms of accuracy and F1-score. In particular, F1 scores improve by up to 20\% in both A and D categories. The results suggest that the improvement in performance can be attributed to MR-DAN's superiority on capturing complex structure information in dialogue and flexibility of our class-wise adaptive SSL.  Additionally, and in line with expectations, the multimodal large models demonstrate superior performance compared to the text-based models.

\paragraph{Main result on MIntRec2.0}
To further demonstrate the improvement of our proposed framework in dialogue modeling ability, we test it on the publicly available dataset MultiRec2.0. The experimental results from Table\ref{tab:performance_comparison} show that our DialogGraph-LLM architecture achieves state-of-the-art performance on the MIntRec 2.0 benchmark for dialogue intent recognition, for both in-scope and out-of-scope evaluations, which indicates the robustness of our approach to explicitly disentangle and model diverse utterance relationships, and identify intricate intents in real-world scenarios. 
The improvement in out-of-scope performance can also be attributed to the synergy between our novel MR-DAN and the sophisticated classification capabilities in the adopted LLM. 

\subsection{Ablation Studies}
To comprehensively evaluate the contributions of various components of our DialogGraph-LLM, we perform ablation studies on its key components and present the result in Table~\ref{tab:ablation_study}.

\begin{table}[htbp] 
\centering
\captionsetup{font=small, labelfont=bf, width=1.5\textwidth}
\caption{Ablation studies of DialogGraph-LLM}
\label{tab:ablation_study}
\begin{tabular}{@{}lcccc@{}} 
\toprule
\multirow{2}{*}{Methods} & \multicolumn{2}{c}{MarketCalls} & \multicolumn{2}{c}{MIntRec2.0 (IS+OS)} \\
\cmidrule(lr){2-3} \cmidrule(lr){4-5} 
 & Acc(\%) & F1(\%) & Acc(\%) & F1(\%) \\
\midrule
\textbf{DialogGraph-LLM } & \textbf{77.31} & \textbf{76.82} & \textbf{64.32} & \textbf{58.14} \\
w/o MR-DAN                & 68.38          & 67.69          & 61.83          & 55.86          \\
w/ Global Threshold       & 73.58          & 72.93          & /              & /              \\
\bottomrule
\end{tabular}
\end{table}

\paragraph{Effectiveness of MR-DAN}
Removing MR-DAN negatively impacted performance on both datasets, with a particularly pronounced effect on MarketCalls.

As shown in Table~\ref{tab:ablation_study}, the absence of  MR-DAN component leads to a significant decline in the overall performance on the \textbf{MarketCalls} dataset, dropping Accuracy by 8.93\% and F1 score by 9.13\%. This highlights the critical role of MR-DAN for this dataset. The purchase intent recognition task in MarketCalls heavily relies on accurately tracking subtle changes, speaker dynamics, and contextual evolution throughout the dialogue. MR-DAN explicitly models temporal, speaker, and semantic relationships, providing crucial structural information that the LLM backbone alone struggles to adequately capture, especially in low-data regimes.

On the \textbf{MIntRec2.0} dataset, removing MR-DAN similarly results in performance degradation, with Accuracy falling from  64.32\% to 61.83\% and F1 dropping from 58.14\% to 55.86\%. The intent categories in this dataset are inherently less dependent on long-distance, complex structures, so the contribution of MR-DAN is not as significant as it is in MarketCalls. But the impacts can still further validate the generalizability and effectiveness of MR-DAN in capturing and utilizing dialogue structural information to improve intent recognition accuracy across different conversational scenarios.

\paragraph{Effectiveness of Adaptive SSL Strategy}
To evaluate the effectiveness of our adaptive SSL strategy featuring class-specific thresholds, we replace the class-specific adaptive thresholds with a single global threshold. This leads to reduction in Accuracy and F1 score on MarketCalls by 3.73\% and 3.89\%, respectively. This confirms the effectiveness of the class-adaptive strategy, which further improves over fixed threshold semi-supervised learning on imbalanced data by more flexibly accommodating class-wise confidence distributions. This strategy, designed for MarketCalls given its scarcity in labels, was not applied to MIntRec2.0.

\subsection{Sensitivity Studies}
To evaluate the influence of key hyperparameters in MR-DAN on overall performance, we conduct sensitivity analyses on the MarketCalls dataset and MIntRec2.0 dataset.

\setlength{\belowcaptionskip}{14pt}
\begin{figure}[htbp]
      \centering
      \includegraphics[scale=0.2]{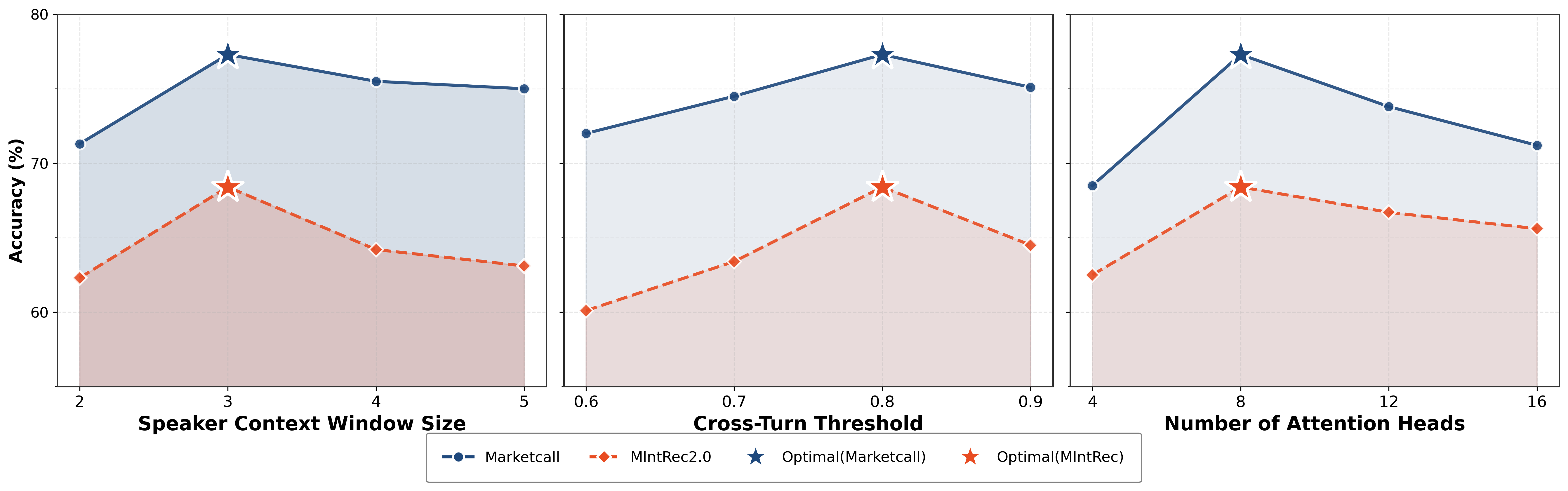}
      \caption{Sensitivity analysis1
      1of model accuracy to key MR-DAN hyperparameters}
      \label{fig_sensitivity}
\end{figure}

\noindent\textbf{Speaker Context Window.}\quad  This parameter dictates the number of preceding same-speaker utterances linked via Speaker Edges, controlling the scope of explicit speaker history modeling. Testing $k \in \{2, 3, 4, 5\}$, we find optimal performance at $k=3$ (77.3\% accuracy). This suggests linking the three most recent utterances effectively captures relevant speaker-specific context. Smaller values ($k=2$) likely miss crucial context, while larger values ($k=4, 5$) introduce noise from temporally distant, less relevant utterances.


\noindent\textbf{Cross-Turn Similarity Threshold.}\quad $\theta$ sets the starting threshold for establishing Cross-Turn Edges based on semantic similarity, which is subsequently adjusted dynamically during training. Evaluating initial values from 0.6 to 0.9, accuracy peaked at $\theta=0.8$. Lower initial values (0.6, 0.7) may introduce excessive noisy edges that hinder the dynamic refinement process, while a higher initial value (0.9) risks prematurely pruning potentially useful semantic connections.

\noindent\textbf{Number of Attention Heads.}\quad $H$ determines the capacity for parallel multi-relational modeling across MR-DAN's different edge types. We tested $H \in \{4, 8, 12, 16\}$. Performance improved substantially from $H=4$ to peak at $H=8$, indicating that 4 heads provide insufficient capacity to model the diverse relational inter-dependencies effectively. Increasing heads further ($H=12, 16$) led to diminishing returns and slightly reduced accuracy, potentially due to overfitting or excessive fragmentation of relational information.

\section{Conclusion}\label{sec:conclusion}
In this work, we introduce DialogGraph-LLM for end-to-end audio dialogue intent recognition, leveraging a novel MR-DAN module to encode conversational structure, and integrating these representations with audio features within an LLM framework using structured prompting. 
To address data scarcity, we develop an adaptive semi-supervised learning strategy, leading to significant improvements on both the MarketCalls and MIntRec 2.0 datasets. 

We select Qwen2.5-Omni-7B as our LLM backbone in the current work for all experiments. While our proposed DialogGraph-LLM framework can accomodate other LLM backbones, we do not experiment with different LLM options. Meanwhile, our proposed MR-DAN makes predefined structural assumptions, and our SSL strategy may incur noise propagation from pseudo-labels. Future work will focus on exploring a diverse LLM backbone options, more flexible graph representations, and more robustness SSL strategies.
\bibliography{main}


\end{document}